\documentclass[iop]{emulateapj}
\usepackage{color}
\usepackage{graphicx}
\usepackage{amsmath}
\usepackage{color}
\usepackage[colorlinks,linkcolor=black,anchorcolor=black,citecolor=blue,hyperfootnotes=true]{hyperref}
\usepackage[section]{placeins}

\usepackage{amssymb}
\usepackage{amsmath}
\usepackage{color}

\usepackage{soul} 
\usepackage{array}

\shorttitle{Double compact object detection and lensing }
\shortauthors{Pi{\'o}rkowska-Kurpas  et al.}

\begin{document}

\title{Inspiraling double compact object detection and lensing rate -- forecast for DECIGO and B-DECIGO}

\author{Aleksandra Pi{\'o}rkowska-Kurpas\altaffilmark{1,2}, Shaoqi Hou\altaffilmark{3}, Marek Biesiada\altaffilmark{1,4}, Xuheng Ding\altaffilmark{3,5}, Shuo Cao\altaffilmark{1$\ast$}, Xilong Fan\altaffilmark{3$\dag$}, Seiji Kawamura\altaffilmark{6}
and Zong-Hong Zhu\altaffilmark{1,3}}

\altaffiltext{1}{Department of Astronomy, Beijing Normal University,
Beijing 100875, China; \emph{zhuzh@bnu.edu.cn; caoshuo@bnu.edu.cn}}
\altaffiltext{2}{Institute of Physics, University of Silesia, 75
Pu{\l}ku Piechoty 1, 41-500 Chorz{\'o}w, Poland}
\altaffiltext{3}{School of Physics and Technology, Wuhan University,
Wuhan 430072, China} \altaffiltext{4}{National Centre for Nuclear
Research, Pasteura 7, 02-093 Warsaw, Poland;
\emph{Marek.Biesiada@ncbj.gov.pl}} \altaffiltext{5}{Department of
Physics and Astronomy, University of California, Los Angeles, CA
90095-1547, USA } \altaffiltext{6}{Department of Physics, Nagoya
University, Nagoya, Aichi 464-8602, Japan}

\begin{abstract}
Emergence of gravitational wave (GW) astronomy revived the interest
in exploring the low frequency GW spectrum inaccessible from the
ground. Satellite GW observatory DECIGO in its original
configuration and the currently proposed smaller scale B-DECIGO are
aimed to cover deci-Hertz part of the GW spectrum, which fills the
gap between LISA mili-Hertz and deca- to kilo-Hertz range probed by
ground-based detectors. In this paper we forecast the detection
rates of inspiraling double compact objects (DCOs) and the
unresolved confusion noise from these sources in DECIGO and
B-DECIGO. In the context of DECIGO we use, for the first time, the
population synthesis intrinsic inspiral rates of NS-NS, BH-NS and
BH-BH systems. We also estimate the expected gravitational lensing
rates of such sources for DECIGO and B-DECIGO. The result is that
yearly detection of resolvable DCOs inspirals for the DECIGO is of order of ${\cal O}(10^2)$ for NS-NS, ${\cal O}(10^3)$ for BH-NS and ${\cal O}(10^5)$ for BH-BH systems, while for a much smaller scale B-DECIGO they are
about ${\cal O}(10)$ for NS-NS, ${\cal O}(10^2)$ for BH-NS and ${\cal O}(10^5)$ for BH-BH systems.
Taking into account that considerable part of these events would be detectable by ground-based GW observatories the significance of
DECIGO/B-DECIGO could be substantial. 
Due to contamination by unresolved sources, both DECIGO and B-DECIGO will not be able to register lensed NS-NS or BH-NS systems, but the lensed BH-BH systems could be observed at the rate of about 50 per year in DECIGO. Smaller scale B-DECIGO will be able to detect a few lensed BH-BH systems per year. We also
address the question of the magnification bias in the GW event catalogs of DECIGO and B-DECIGO.

\end{abstract}

\maketitle

\section{Introduction}

First laboratory detections of gravitational waves (GWs) on Earth
\citep{LIGO_1st} opened up a new branch of science -- GW astronomy.
Continuing efforts of LIGO/Virgo team (with now completed O1, O2 and O3 scientific runs ) brought numerous detections of binary
black hole (BH-BH) mergers \citep{LIGO_run2}, probably the first
mixed black hole - neutron star (BH-NS) merger \cite{BHNS} and the
first detection of binary neutron star (NS-NS) coalescence
\citep{NSNS_merger}.  The NS-NS merger was accompanied by
identification of its electromagnetic (EM) counterpart and its
afterglow was followed up at different EM wavelengths
\citep{Goldstein, Coulter}. This observation moved multimessenger
astronomy to the next level. Besides the tests of general relativity
and modified gravity theories \citep{TestGR}, strong bounds on the
speed of GWs \citep{GWspeed}, GW astronomy has proven that (almost)
all classes of double compact objects (DCOs): NS-NS, BH-NS, BH-BH
really exist in Nature. The event GW190425 is marginally compatible
with NS-NS merger, so the existence of BH-NS systems still needs to
be empirically proven. It is likely that this will happen soon.

Successful operation of ground-based interferometric detectors
revived the interest in broadening the GW spectrum to lower
frequencies (lower than 1 Hz) fundamentally inaccessible from the
ground due to irremovable seismic noise. In particular, LISA was
proposed \citep{LISA2017} as a new generation space mission with a
robust strain sensitivity level at frequencies between 0.1 mHz and
100 mHz over a science lifetime of at least 4 years. The technology
behind LISA, an ESA-led mission expected to be launched by 2034, has
been recently tested by the LISA Pathfinder experiment with
outstanding results \citep{Armano2016}. The remaining decihertz band
is the target of the DECihertz Interferometer Gravitational wave
Observatory (DECIGO), a planned Japanese space-borne GW detector
\citep{Kawamura2019}. Original DECIGO \citep{Seto2001} or currently
proposed smaller-scale version B-DECIGO \citep{Sato2017} are aimed
to cover the low-frequency band extending from the mHz to 100 Hz
range. DECIGO would be able to detect inspiraling DCO --- main
targets to LIGO/Virgo, KAGRA or next generation ET, long time (weeks
to years) before they enter the hectohertz band accessible from the
ground. Moreover, the overlap with ground based GW detectors
sensitivity bands is very advantageous since the joint detection
with DECIGO and ground based detectors (e.g. ET) would greatly
improve the parameter estimation of the binaries.

Reach of the DECIGO will be considerably higher than the next
generation of ground based interferometric detectors, like the ET.
With much higher volume probed, one may expect that non-negligible
number of GW signals from coalescing DCOs would be gravitationally
lensed. Gravitational lensing statistics and magnification
cross-sections by galaxies has been discussed e.g. in
\citep{Zhu1997,Zhu1998}. Rates of GW lensing for LIGO/Virgo have
been calculated in \citet{Tjonnie} and for the ET in
\citet{JCAP_ET1, JCAP_ET2, JCAP_ET3, Mao, Oguri, Yang2019},
respectively. The robust prediction is that the third generation of
GW interferometric detectors would yield 50 - 100 lensed GW events
per year. Strongly lensed GW signals, observed together with their
electromagnetic (EM) counterparts, have been demonstrated to enhance
our understanding regarding fundamental physics \citep{Fan2017,
Collett2017,Cao2019}, dark matter \citep{Liao2018}, and cosmology
\citep{Sereno2010, Sereno2011, Taylor2012, NatureComm, Wei2017}.
Gravitational lensing of GWs has been widely discussed concerning
diffraction effects in lensed GW events \citep{Liao2019,
Takahashi2003, Nakamura1998}, the waveform distortion caused by the
gravitational lensing \citep{Cao2014}, the influence on the
statistical signatures of black hole mergers \citep{Dai2017}.

In this paper we make predictions concerning detection rates of inspiraling DCO systems, confusion noise regarding unresolved DCO systems and finally  gravitational lensing
of GW signals detectable by the DECIGO.

\section{GW background noise from DCO systems}

We start with the assessment of the background noise created by
unresolved DCO systems. In next sections we will discuss the
detection rate of such inspiraling systems and their lensing rate by
background galaxies. In all these considerations one needs the
detector's sensitivity curve and intrinsic DCO merger rates as
input. We will discuss these issues below.

\subsection{DECIGO sensitivity}

DECIGO will be composed of four units of detectors. Each unit is
planned to contain three drag-free spacecrafts to form a nearly
regular triangle. These four units will be tilted $60^{\circ}$
inwards relative to the ecliptic plane to keep its arm lengths
nearly constant, and move around the sun with the orbital period of
1 yr. Centers of the triangular configuration of the units will form
an equilateral triangle. The fourth one will be anchored to one of
the units rotated $180^{\circ}$ to form a Star of David
configuration. Details concerning the DECIGO design could be found
in \citet{Yagi2011}

For the reference design parameters of DECIGO, one can prove
\citep{Yagi2011} that a single triangular detector unit is
equivalent to two L-shaped interferometers rotated by  $45^\circ$.
Their noises are uncorrelated and the noise spectrum of such an
effective L-shaped DECIGO is given by:
\begin{align} \label{sensitivity_DECIGO}
S_h(f) =& 10^{-48} \times \Bigg[ 7.05 \left( 1 + \frac{f^2}{f_p^2} \right) + 4.80\times 10^{-3}\times   \nonumber \\
& \frac{f^{-4}}{1 + \left( \frac{f}{f_p} \right)^2} + 5.33 \times 10^{-4} f^{-4} \Bigg]\;\; Hz^{-1}
\end{align}
where $f_p = 7.36\; Hz$.

The original DECIGO mission concept was proposed in 2001  by
\cite{Seto2001}. What now seems more realistic to be commissioned in
the near future is a scaled smaller project called B-DECIGO. It will
consist of three satellites in a 100 km equilateral triangle, having
sun-synchronous dusk-dawn circular orbits 2000 km above the Earth
\cite{Sato2017, Nakamura2016}. With B-DECIGO operating, we will soon
probe the decihertz window for the first time, completing the full
gravitational spectrum. Therefore we extend our predictions to the
B-DECIGO. We use the noise power spectrum density $S_h(f)$ for
B-DECIGO proposed by \cite{Nakamura2016, Isoyama2018}:
\begin{align} \label{sensitivity_B-DECIGO}
S_h(f) =& 10^{-46} \times [ 4.040 + 6.399 \times 10^{-2} f^{-4}   \nonumber \\
& + 6.399 \times 10^{-3} f^{2} ]\;\; Hz^{-1}
\end{align}

\subsection{DCO merger rate from the population synthesis}

Previous predictions concerning DECIGO \citep{Yagi2011, Yagi2013}
used an analytical approximation of the NS-NS merger rate as a
function of redshift up to redshift $z=5$. Moreover, they did not
assess the BH-BH or BH-NS rates precisely. In this paper, we use the
values of the intrinsic inspiral rates ${\dot n_0(z_s)}$  forecasted
by \citet{BelczynskiII} for each type of DCO at redshift slices
spanning the range of $z \in [0.04;17]$. These data, available  at
{\tt https://www.syntheticuniverse.org} have been used in our previous
papers \citep{JCAP_ET2, JCAP_ET3}, where one can find more details
about them. For the current purpose it will be sufficient to recall
the following facts. 
To account for the varied chemical composition of the
Universe, they performed the cosmological calculations for
two scenarios of galactic metallicity evolution, called
``low–end'' and ``high–end'', respectively. Essentially, they modeled metallicity 
(in fact oxygen abundance, assumed to be correlated with metallicity) by empirical function of galaxy mass (derived from Schechter type distribution). The normalization factor of this relation is redshift dependent, which can be heuristically modeled by a relation, whose coefficient choice (in fact, it's a bit more complicated - see \cite{BelczynskiII}) lead to two distinct metallicity evolution profiles. The first one called ``high-end'' predicts a median
value of metallicity of $1.5\; Z_{\odot}$ at $z \sim 0$. Another one, called ``low-end'' yields a median metallicity value of $0.8 \; Z_{\odot}$. 
Concerning binary system evolution from the
ZAMS to the final formation of DCO binary system (NS-NS, NS-BH or
BH-BH) we will consider four scenarios: standard, optimistic common
envelope, delayed SN explosion and high BH natal kicks as specified
in \citet{BelczynskiII}, where the reader is referred to for more
details. 
In previous papers 
cited above, the median masses: $1.2M_\odot$ for NS-NS, $3.2M_\odot$ for BH-NS, and $6.7M_\odot$ for BH-BH were used according to \cite{BelczynskiI}. However, these were values obtained under the assumption of solar metallicity of initial binary systems. Such scenario was a right guess before the first detections of GWs. Now, the data collected with the LIGO/Virgo detectors 
 demonstrated that observed chirp masses (in particular of BH-BH systems) are much higher. Hence one is forced to change the aforementioned assumption. Moreover, as we will see further on, probability density of DCO inspirals (observable by DECIGO) peak at z=2, when the metallicity was significantly sub-solar.
  Therefore, guided by the real data gathered so far we will adopt different values --  according to \cite{LIGO_run2}. We will assume the median 
 value of BH-BH systems chirp masses reported in their Table III. Since the data on BH-NS systems is more scarce, we will take the value of \cite{Abbott:2020khf}. Hence, the following chirp masses will be adopted by us as representative of DCO inspiraling systems: 
 $1.2\;M_\odot$ for NS-NS, $6.09\; M_\odot$ for BH-NS, and
$24.5\;M_\odot$ for BH-BH.
In order to comply with the assumptions
underlying population synthesis simulation, we assume flat
$\Lambda$CDM cosmology with $H_0 = 70 \; km\,s^{-1}\,Mpc^{-1}$ and $\Omega_m=0.3$.

\subsection{GW background from unresolved DCO systems}

Magnitude of a stochastic GW background is usually characterized by
its fractional energy density per logarithmic frequency interval:
\begin{equation}
\Omega_{GW} = \frac{1}{\rho_{cr}} \frac{d \rho_{GW}}{d \ln{f}}
\end{equation}
where $\rho_{cr} = 3 H_0^2 / 8 \pi G$ is the critical energy density of the Universe. Let us note, that in the context of a confusion noise due to unresolved sources, the distinction between background and foreground is not clear-cut. We will call it background.

According to \cite{Phinney} one can conveniently calculate the energy density parameter
$\Omega_{GW}^{DCO}$ corresponding to the unresolved signals from the DCO systems, as
\begin{equation} \label{Phinney}
\Omega_{GW}^{DCO} =  \frac{8 \pi^{5/3}}{9c^2 H_0^2} (G {\cal M})^{5/3} f^{2/3} \int_0^\infty \frac{\dot{n}(z)}{(1+z)^{4/3} H(z)} dz
\end{equation}
where ${\cal M}$ is the chirp mass of the DCO system (i.e. NS-NS,
BH-NS or BH-BH binary), ${\dot n}(z)$ is the DCO merger rate per
proper time per comoving volume at redshift $z$, and the Hubble
parameter $H(z)$ is given by $H(z)^2 = H_0^2 [ \Omega_m (1+z)^3 +1 -
\Omega_m ] $. Calculating numerically the integral in
Eq.(\ref{Phinney}) one obtains:
\begin{equation} \label{OmegaGW}
    \begin{split}
\Omega_{GW}^{DCO} =& \Omega_{0}^{DCO} \left( \frac{H_0}{70\;km s^{-1} Mpc^{-1}} \right)^{-3} \times\\
&\left( \frac{{\cal M}}{{\cal M}_{DCO}} \right)^{5/3} \left( \frac{f}{1 Hz} \right)^{2/3}
    \end{split}
\end{equation}
where ${\cal M}_{DCO}$ is the median value of the DCO considered,
i.e. $({\cal M}_{NSNS}; {\cal M}_{BHNS}; {\cal M}_{BHBH}) = (1.22;  6.09; 24.5)\;M_{\odot}$ and values of $\Omega_{0}^{DCO}$ coefficients
are reported in Table~\ref{Omega_GW}.

In order to compare with the detector's noise power spectrum the
normalized energy density $\Omega_{GW}^{DCO}$ should be expressed as
the total sky-averaged GW 
spectrum:
\begin{equation}
S_h^{GW,DCO} = \frac{4}{\pi} f^{-3} \rho_{cr} \Omega_{GW}^{DCO}.
\end{equation}

\begin{figure}[h]
\begin{center}
\includegraphics[width=0.99\columnwidth]{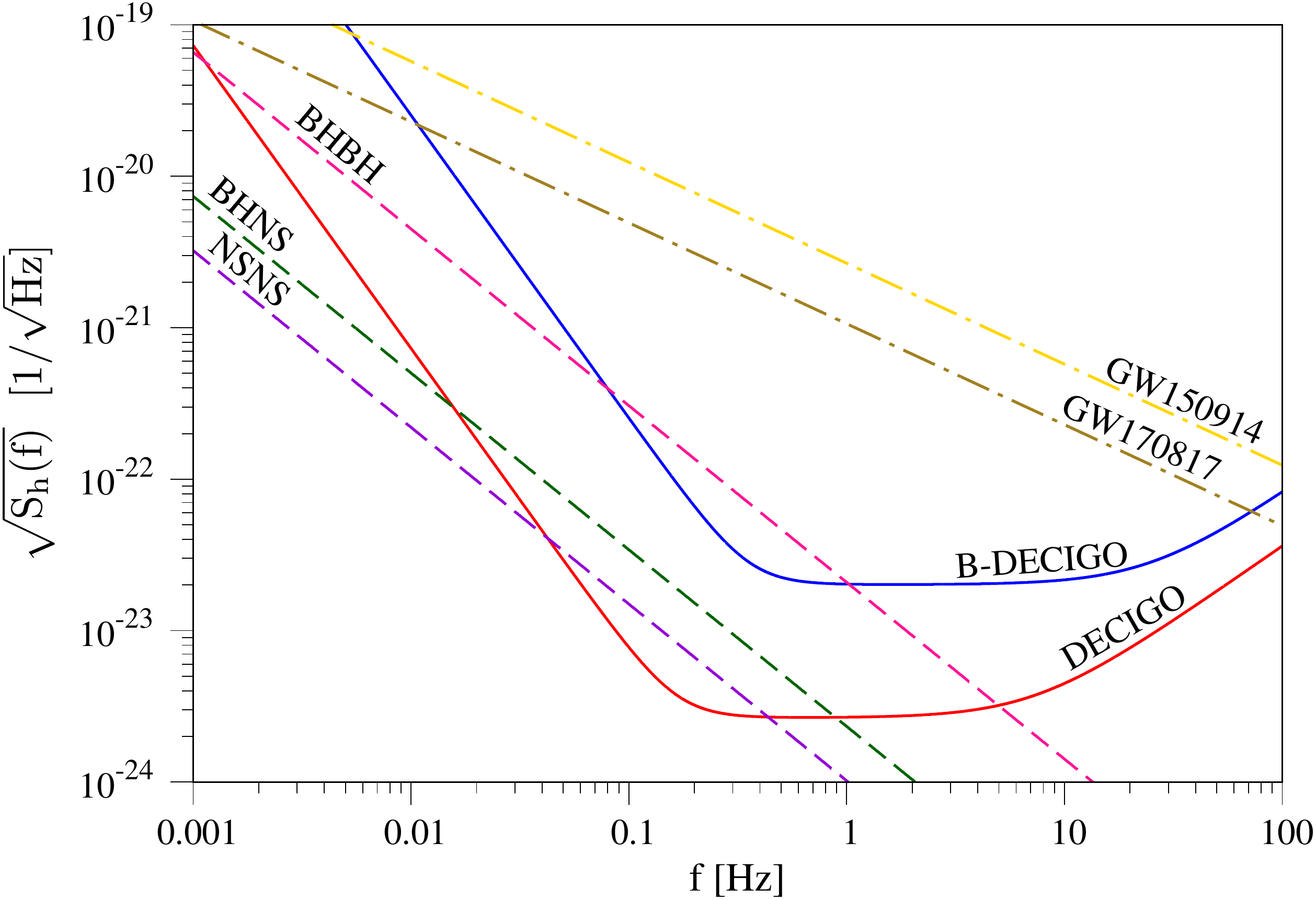}
\end{center}
\caption{Confusion noise due to DCO systems (NS-NS, BH-NS and BH-BH) superimposed on the noise spectrum density for DECIGO and B-DECIGO. Inspiral rates correspond to the standard scenario and 
``low-end'' metallicity evolution.
The effective squared spectrum density corresponding to GW150914 and GW170817 events has been also included.
\label{confusion_noise}}
\end{figure}

The background spectrum of three DCO populations imposed on the
DECIGO sensitivity curve is shown in Figure~\ref{confusion_noise}.
Let us note that B-DECIGO will be much less contaminated from the
unresolved DCO systems, yet the events like those detected by
LIGO/Virgo will be detectable enabling their discovery and study
long before they will enter the ground-based detectors  sensitivity band. 
As discussed in details in \cite{Isoyama2018}, time to coalescence can be estimated as: $t_c = 1.03 \times 10^6 \; {\mathrm{s}} \; ({\cal M}_z /30.1 \; M_{\odot})^{-5/3} (f / 0.1 \; \mathrm{Hz})^{-8/3} $,  which means that GW150914 and GW170817 could have been visible in (B-)DECIGO band for $\sim 10$ days and $\sim  7$ yrs prior to coalescence with large numbers of GW cycles ($10^5$ and $10^7$ , respectively).

\section{Detection rate for unlensed events} \label{sec:rates}

Matched filtering signal-to-noise ratio 
 for a single detector reads \citep{Taylor2012}:
 \begin{equation} \label{SNR}
 \rho = 8 \Theta \frac{r_0}{d_L(z_s)} \left(\frac{{\cal M}_z}{1.2\;M_{\odot}}\right)^{5/6}
 \sqrt{\zeta(f_{max})}
 \end{equation}
where $d_L$ is the luminosity distance to the source, $\Theta$ is
the orientation factor capturing part of sensitivity pattern due to
(usually non-optimal) random relative orientation of DCO system with
respect to the detector (more details below). Zeta parameter
capturing the overlap between the signal and the detector's
sensitivity band width is defined as:
\begin{equation} \label{zeta}
\zeta(f_{max}) = \frac{1}{x_{7/3}} \int_0^{2f_{max}}\frac{ df\; (\pi M_{\odot})^2}{(\pi f M_{\odot})^{7/3} S_h(f)} \nonumber
\end{equation}
where $2f_{max}$ is the wave frequency at which the inspiral detection template ends and $x_{7/3} = (\pi M_{\odot} )^{-1/3} f_{7/3}$ (see below for definiton of $f_{7/3}$). Calculating $\zeta$ factor with the above definition one obtains  $\zeta(f_{max}) \approx 1$ for all DCO systems considered. However, the confusion noise of unresolved systems will influence our ability to detect inspiraling DCO systems. Therefore, we should use the modified noise spectrum $S_h(f) + S_h^{GW,DCO}(f)$ (system by system, and scenario by scenario). There are two ways to do that, both leading to the same results. First, is the take the detector's characteristic distance $r_0$ (defined below) and hence $x_{7/3}$ calculated from the detector's noise $S_h(f)$ but modified noise in the calculation of $\zeta(f_{max})$ (in $x_{7/3}$ noise is not modified).  In this case $\zeta$ factors are considerably smaller. Second approach is to consider noise spectrum modified by DCO confusion noise and regard different detector's characteristic distances $r_0$ (system by system and scenario by scenario). Then  $\zeta(f_{max}) \approx 1$ is valid. 
%
%
By $r_0$ we denote the detector's characteristic distance parameter,
which can be estimated according to:
 \begin{equation} \label{r0_new}
 r_0^2 = \frac{5}{192 \pi^{4/3}} \left( \frac{3}{20} \right)^{5/3} \frac{( G M_{\odot})^{5/3}}{c^3} f_{7/3} \nonumber
\end{equation}
where:
\begin{equation} \label{f7/3}
 f_{7/3} = \int_0^{\infty}  \left[ f^{7/3} S_h(f) \right]^{-1}\; df
 \nonumber
\end{equation}

Using DECIGO sensitivity Eq.(\ref{sensitivity_DECIGO}) one gets:
$$r_0 = 6709 \; Mpc. $$ Similarly, from the B-DECIGO sensitivity
given by Eq.(\ref{sensitivity_B-DECIGO}) one obtains $r_0 = 535 \;
Mpc. $, meaning that B-DECIGO will be able to probe about a 1000
times smaller volume than DECIGO.

The orientation factor $\Theta$ is defined as
 \begin{equation} \label{Theta}
 \Theta = 2 [ F_{+}^2(1 + \cos^2{\iota} )^2 + 4 F_{\times}^2 \cos^2{\iota} ]^{1/2}
 \end{equation}
where $F_{+} = \frac{1}{2} (1 + \cos^2{\theta}) \cos{2\phi} \cos{2 \psi} - \cos{\theta} \sin{2 \phi} \sin{ 2 \psi}$ and
$F_{\times} = \frac{1}{2} (1 + \cos^2{\theta}) \cos{2\phi} \sin{2 \psi} + \cos{\theta} \sin{2 \phi} \cos{ 2 \psi}$ are the interferometer strain responses to different polarizations of gravitational wave. The angles $(\theta, \phi)$ describe orientation (polar angles) of direction to the source with respect to the detector plane, $(\psi, \iota)$ are the angles describing DCO orbit  orientation
with respect to the plane tangent to the celestial sphere at source location (so called polarization angle and inclination). The above formulae for $F_{+/ \times}$ are for just one L-shaped detector $F_{I;+/ \times}$, for the second one it would be
$F_{II; +/ \times} = F_{I,+/ \times}(\theta, \phi - \pi/4, \psi, \iota).$

Probability distribution for $\Theta$ calculated under assumption of
uncorrelated orientation angles $(\theta, \phi, \psi, \iota)$ is
known to be of the following form \citep{Finn96}:
\begin{eqnarray} \label{P_theta}
P_{\Theta}(\Theta) &=& 5 \Theta (4 - \Theta)^3 /256, \qquad {\rm if}\;\;\;
0< \Theta < 4  \\
P_{\Theta}(\Theta) &=& 0, \qquad {\rm otherwise} \nonumber
\end{eqnarray}

The yearly detection rate of DCO sources originating at redshift
$z_s$ and producing the signal with SNR exceeding the detector's
threshold $\rho_0=8$ (assumption made in previous DECIGO studies
\citep{Yagi2011, Isoyama2018}) can be expressed as:
 \begin{equation} \label{Ndot}
{\dot N}(>\rho_0|z_s) = \int_0^{z_s} \frac{d {\dot N}(>\rho_0)}{dz} dz
\end{equation}
where
\begin{equation} \label{rate_nl}
\frac{d {\dot N}(>\rho_0)}{dz_s} = 4\pi \left( \frac{c}{H_0} \right)^3 \frac{{\dot n_0}(z_s)}{1+z_s}\;  \frac{{\tilde r}^2(z_s)}{E(z_s)} \; C_{\Theta}(x(z_s, \rho_0))
\end{equation}
${\dot n_0}(z_s)$ denotes intrinsic coalescence rate at redshift
$z_s$, $C_{\Theta}(x) = \int_x^{\infty} P_{\Theta}(\Theta) d\Theta$
and $x(z, \rho) = \frac{\rho}{8} (1+z)^{1/6} \frac{c}{H_0}
\frac{{\tilde r}(z)}{r_0} \left( \frac{1.2\; M_{\odot}}{{\cal M}_0}
\right)^{5/6} \zeta(f_{max})^{-1/2}$, i.e. $x(z,\rho)$ is a transformed Eq.~(\ref{SNR})
where $\Theta$ is treated as a variable.

\begin{figure}[tbp]
\centering
\includegraphics[width=0.99\columnwidth]{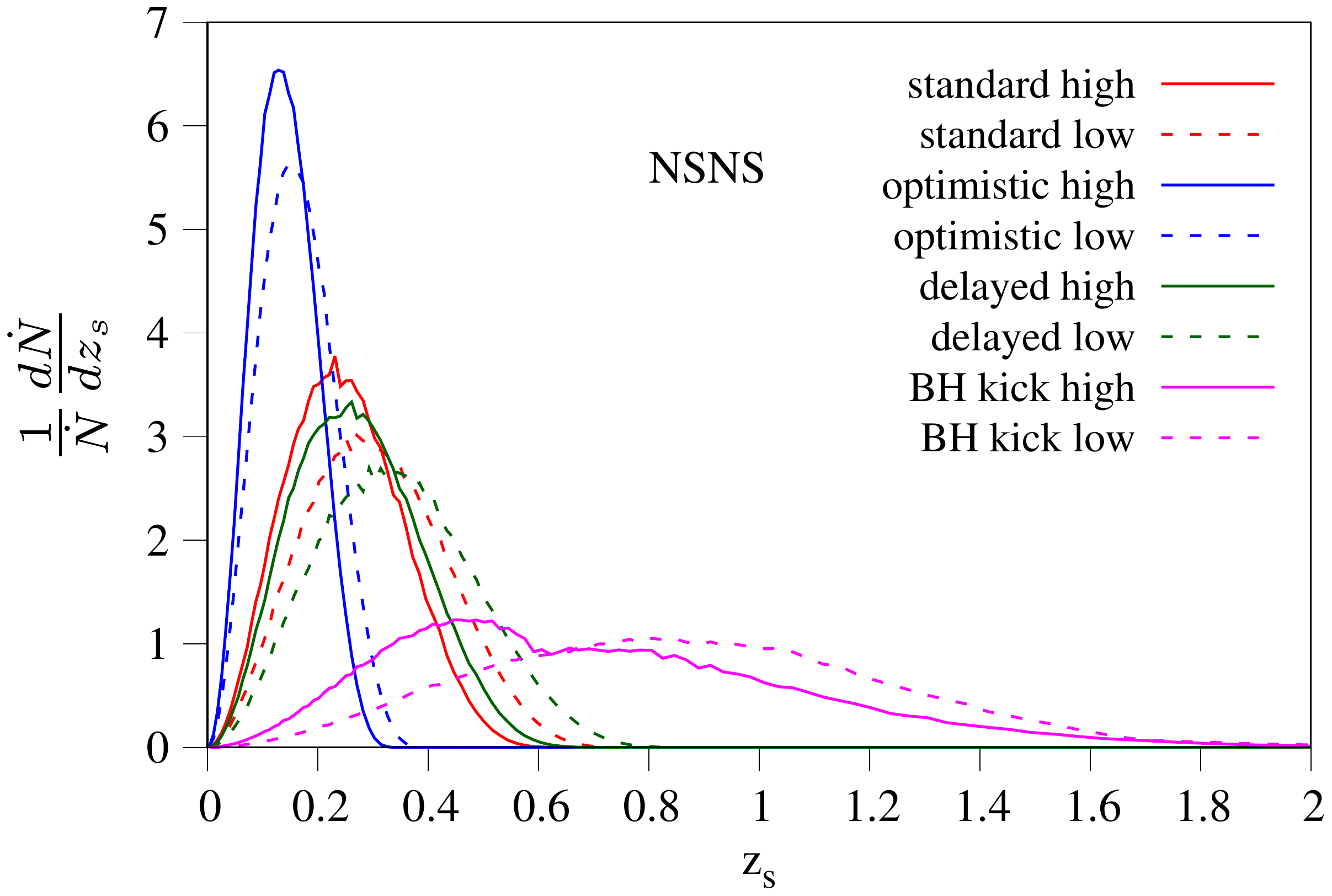}
\includegraphics[width=0.99\columnwidth]{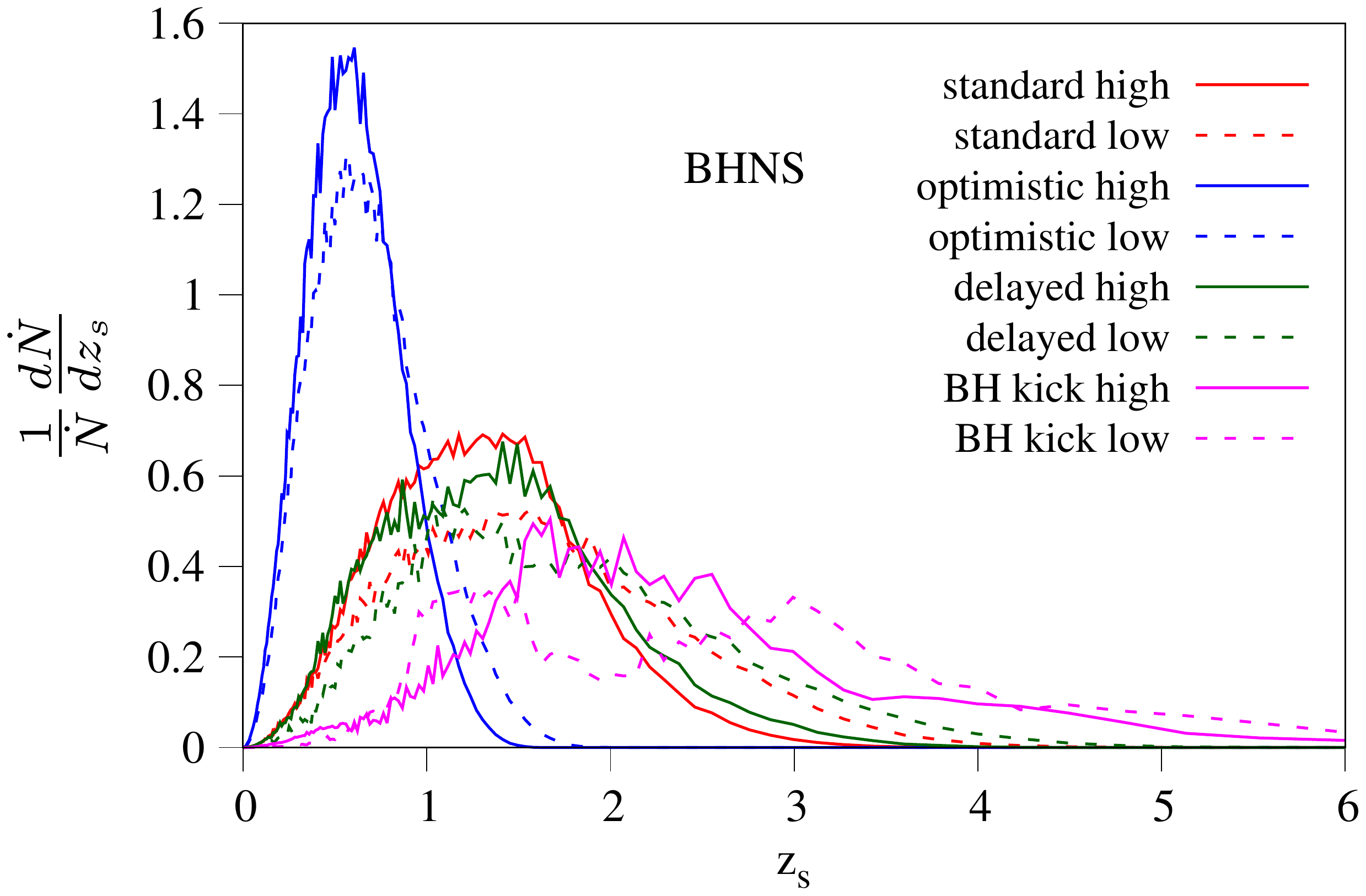}
\includegraphics[width=0.99\columnwidth]{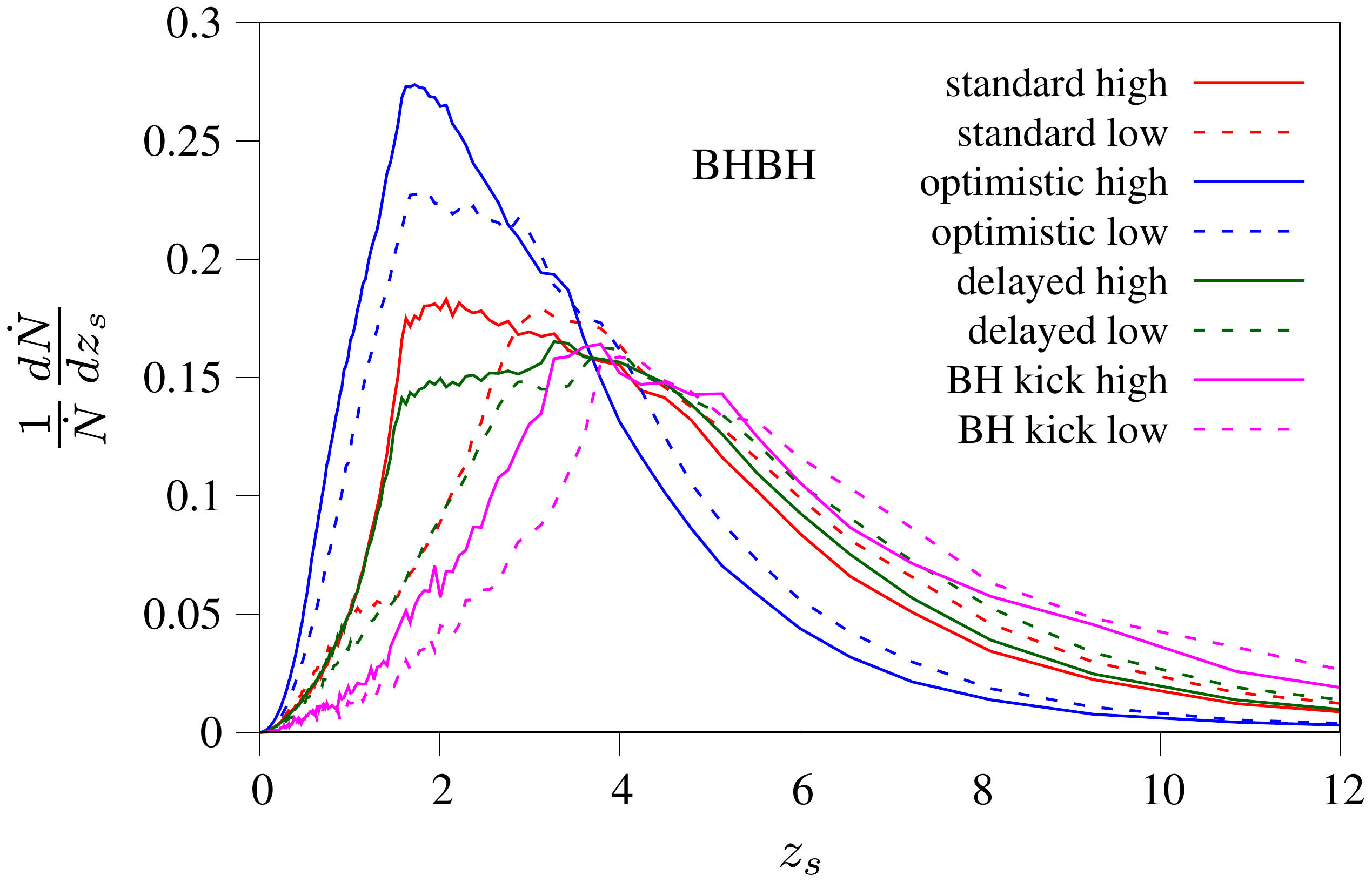}
\caption{Probability density of DCO inspiral events as a function of redshift for the DECIGO. Different colors refer to different scenarios:
red -- standard, blue -- OCE, green -- delayed SN, magenta -- high BH kicks. 
Solid line corresponds to the high-end metallicity evolution, dashed line -- low-end metallicity.
\label{diff rate}}
\end{figure}

Throughout this paper we use the values of inspiral rates ${\dot
n_0}(z_s)$ obtained with {\tt StarTrack} evolutionary code. The
results are summarized in Table~\ref{rates}. Probability density of
DCO inspiral events as a function of redshift is shown in
Fig~\ref{diff rate}. 
One can see that even though B-DECIGO
is considerably smaller scale enterprize, it would be able to
register tens of resolvable NS-NS inspirals and hundred thousands of resolvable BH-BH inspirals per year.

\section{Lensed GW signals statistics}

We assume that the population of lenses comprise only elliptical
galaxies  modeled as singular isothermal spheres (SIS). This
assumption is supported by galaxy strong lensing studies
\citep{Koopmans09}. Einstein radius, which is a characteristic
angular scale of separation between images in the SIS model can be
expressed as: $\theta_E = 4 \pi \left( \frac{\sigma}{c} \right)^2
\frac{d_A(z_l,z_s)}{d_A(z_s)} $, where $\sigma$ is the velocity
dispersion of stars in lensing galaxy, $d_A(z_l,z_s)$ and $d_A(z_s)$
are angular diameter distances between the lens and the source and
to the source, respectively \citep{Cao2012,Cao2015}. Calculations
are simplified if one expresses the angular distance of the image
from the center of the lens $\theta$ and the angular position of the
source $\beta$ as dimensionless parameters: $x =
\frac{\theta}{\theta_E}$, $ y = \frac{\beta}{\theta_E}$. Then the
necessary condition for strong lensing (appearance of multiple
images) is $y<1$. Brighter $I_{+}$ and fainter image $I_{-}$ form at
locations $x_{\pm} = 1 \pm y$ with magnifications: $\mu_{\pm} =
\frac{1}{y} \pm 1$. Gravitationally lensed GW signal would come from
these two images with appropriate relative time delay $\Delta t =
\Delta t_0 (\sigma, z_l, z_s) y$ (for details see \citep{JCAP_ET1,
JCAP_ET2}) and with different amplitudes: $h_{\pm} =
\sqrt{\mu_{\pm}} \; h(t) = \sqrt{\frac{1}{y} \pm 1}\; h(t) $ where
$h(t)$ denotes the intrinsic amplitude (i.e. the one which would
have been observed without lensing). In order to observe lensed
image $I_+$ or $I_-$ the detected SNR $\rho_{\pm}$ must exceed the
threshold for detection $\rho_0 = 8$. This happens, if the
misalignment of the source with respect to the optical axis of the
lens satisfies:
\begin{equation} \label{y_condition}
y_{\pm} \leq y_{\pm,max} = \left[ \left( \frac{8}{\rho_{intr.}} \right)^2 \mp 1 \right]^{-1}
\end{equation}
where $\rho_{intr.}$ is the intrinsic SNR of the (unlensed) source.

\begin{table*}[t]
\caption{Numerical factors $\Omega_0^{DCO}$ in the
Eq.(\ref{OmegaGW}) for different classes of DCO systems under
different evolutionary scenarios, assuming ``low-end'' and
``high-end'' metallicity evolution. } \label{Omega_GW}
\begin{center}  
\begin{tabular}{cccccc}
\hline
\\
 && $\Omega_0^{DCO} \times 10^{-12}$ && \\
 \\
 \hline
 \\
Evolutionary scenario  &  standard & optimistic & delayed SN & high BH kicks  \\

\hline
NS-NS & & & & \\
low-end metallicity & 1.33 & 10.26 & 1.51 & 1.34 \\
high-end metallicity & 2.17 & 9.95 & 2.44 & 2.14 \\

\hline
BH-NS & & & & \\
low-end metallicity & \color{black}{ 6.94} & \color{black}{ 18.79} & \color{black}{ 3.74} & \color{black}{ 0.68} \\
high-end metallicity & \color{black}{ 4.10} & \color{black}{ 14.30} & \color{black}{ 2.22} & \color{black}{ 0.48} \\

\hline
BH-BH & & & & \\
low-end metallicity & \color{black}{ 554.58} & \color{black}{ 2983.01} & \color{black}{ 427.78}  & \color{black}{ 20.88} \\
high-end metallicity & \color{black}{ 368.68} & \color{black}{ 2026.12} & \color{black}{ 262.41} & \color{black}{ 13.65} \\

\hline
\end{tabular}
\end{center}
\end{table*}

\begin{table*}[tbp]
\caption{Yearly detection rate of inspiraling DCOs of different
classes under different evolutionary scenarios, assuming ``low-end''
and ``high-end'' metallicity evolution. Predictions for the DECIGO \color{black}{and B-DECIGO}. Yearly detection rates of resolvable DCO systems are reported.
} \label{rates}
\begin{center}  
\begin{tabular}{cccccc}

\hline
\\
 && \color{black}{Yearly detection rate for DECIGO} && \\
 \\
\hline

\hline
\\
\color{black}{Evolutionary scenario} & standard & optimistic CE & delayed SN & high BH kicks \\
\\
\hline

 NS-NS &&&&& \\
low-end metallicity & $ 233.1$ & $ 119.$ & $ 335.5$ & $3054.4$ \\
high-end metallicity & $439.6$ & $ 203.9$ & $ 707.3$ & $ 8807.7$ \\
\hline

 BH-NS &&&&& \\
low-end metallicity & $2688.9$ & $ 1239.5$ & $ 1838.6$ & $ 1877.6$ \\
high-end metallicity & $ 2000.$ & $ 1314.6$ & $ 1614.5$ & $ 1613.7$ \\

\hline

 BH-BH &&&&& \\
low-end metallicity & $ 207755.2$ & $ 384698.$ & $ 178991.7$ & $ 20125.8$ \\
high-end metallicity & $166436.$ & $ 360001.5$ & $ 145583.5$ & $ 15379.5$ \\

\hline

TOTAL &&&&& \\
low-end metallicity & $ 210677.2$ & $386056.5$ & $ 181165.8$ & $ 25057.8$ \\
high-end metallicity & $ 168875.6$ & $ 361520$ & $ 147905.3$ & $ 25800.9$ \\

\hline
\\
 && \color{black}{Yearly detection rate for B-DECIGO} && \\
 \\
\hline
\\
\color{black}{Evolutionary scenario} & standard & optimistic CE & delayed SN & high BH kicks \\
\\
\hline

 NS-NS &&&&& \\

low-end metallicity & $ 27.9$ & $ 19.5$ & $ 38.1$ & $109.7$ \\
high-end metallicity & $ 54.4$ & $ 42.2$ & $ 78.7$ & $139.$ \\

\hline

 BH-NS &&&&& \\
low-end metallicity & $ 172.6$ & $175.9$ & $ 138.7$ & $ 91.1$ \\
high-end metallicity & $ 182.8$ & $ 267.8$ & $ 142.4$ & $ 90.7$ \\

\hline

BH-BH &&&&& \\
low-end metallicity & $ 54049.4$ & $ 94584.8$ & $ 47625.3$ & $ 7015.$ \\
high-end metallicity & $ 53220.2$ & $ 112221.1$ & $49773.8$ & $ 5078.8$ \\

\hline

TOTAL &&&&& \\
low-end metallicity & $ 54249.9 $ & $ 94780.2$ & $ 47802.1$ & $7215.8$ \\
high-end metallicity & $ 53457.4$ & $ 112531.1$ & $ 49994.9$ & $5308.5$ \\

\hline

\end{tabular}
\end{center}
\end{table*}

Consequently, the cross section for lensing is (see e.g.\citep{JCAP_ET1}):
\begin{equation} \label{cross_section}
S_{cr, \pm}(\sigma, z_l, z_s, \rho) = 16 \pi^3 \left( \frac{\sigma}{c} \right)^4 \left( \frac{{\tilde r}_{ls}} {{\tilde r}_{s}} \right)^2 y_{\pm, max}^2
\end{equation}
and the optical depth for lensing leading to magnifications of
$I_{+}$ and $I_{-}$ images above the threshold reads:
\begin{equation} \label{tau}
    \begin{split}
\tau_{\pm}(z_s, \rho) =& \frac{1}{4 \pi} \int_0^{z_s}\; dz_l \; \int^{\infty}_0 \; d \sigma \; 
4 \pi \left( \frac{c}{H_0} \right)^3 \times \\ &\frac{ {\tilde r}_l^2}{E(z_l)} S_{cr, \pm}(\sigma, z_l, z_s) \frac{d n}{d \sigma}.
    \end{split}
\end{equation}

We model the velocity dispersion distribution in the population of
lensing galaxies as a modified Schechter function $\frac{d n}{d
\sigma} = n_{*} \left( \frac{\sigma}{\sigma_{*}} \right)^{\alpha}
\exp{\left( - \left( \frac{\sigma}{\sigma_{*}} \right)^{\beta}
\right)} \frac{\beta}{\Gamma (\frac{\alpha}{\beta}) }
\frac{1}{\sigma}$ with parameters $n_{*}$,$\sigma_{*}$,${\alpha}$
and $\beta$ taken after \cite{Choi2007}. The choice of this
particular model, despite the existence of more recent data on
velocity dispersion distribution functions is motivated by its best
representing the pure elliptical galaxy population in agreement with
our model assumption \citep{JCAP_ET2,Cao2012b}.

\begin{figure}[htbp]
\centering
\includegraphics[width=0.99\columnwidth]{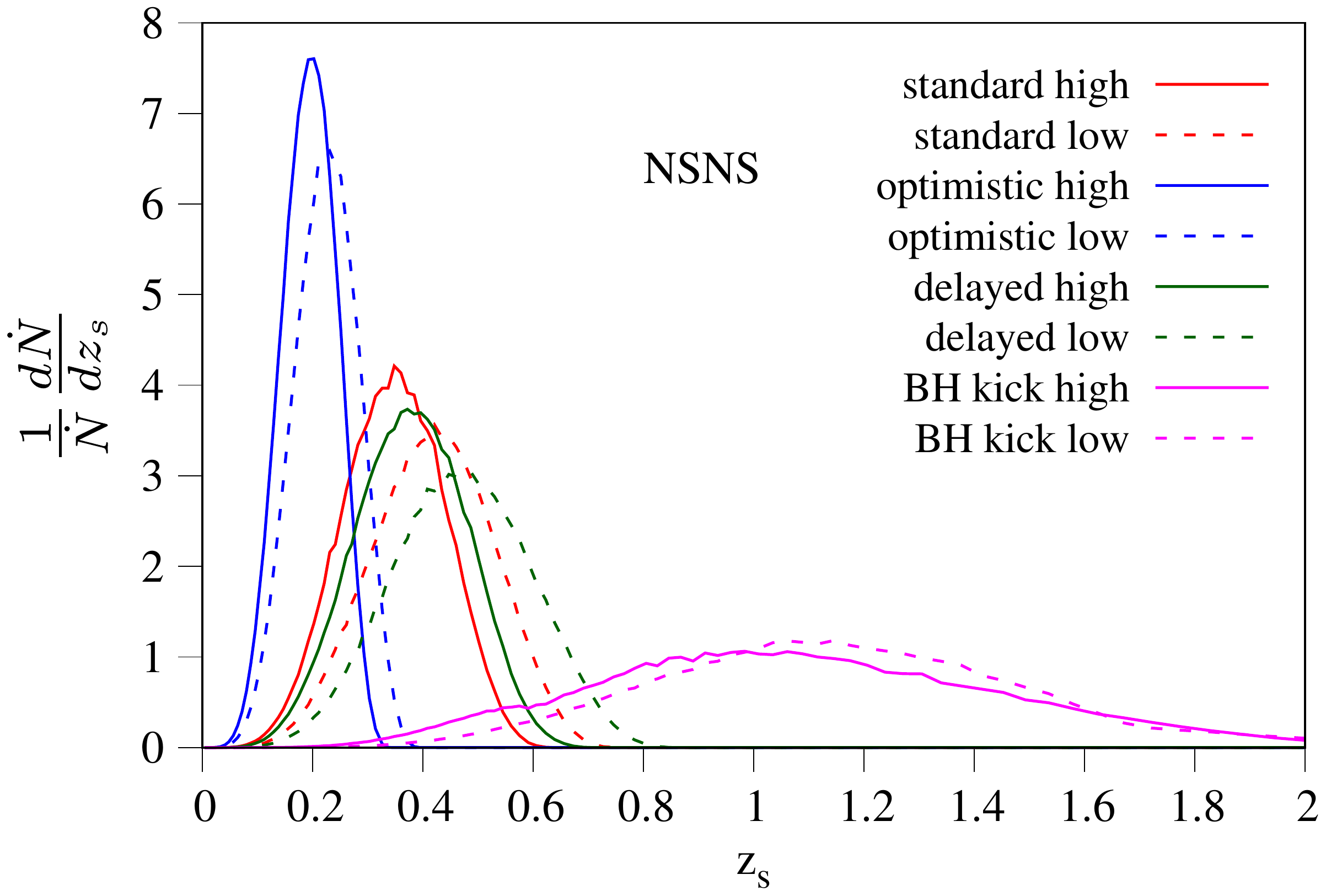}
\includegraphics[width=0.99\columnwidth]{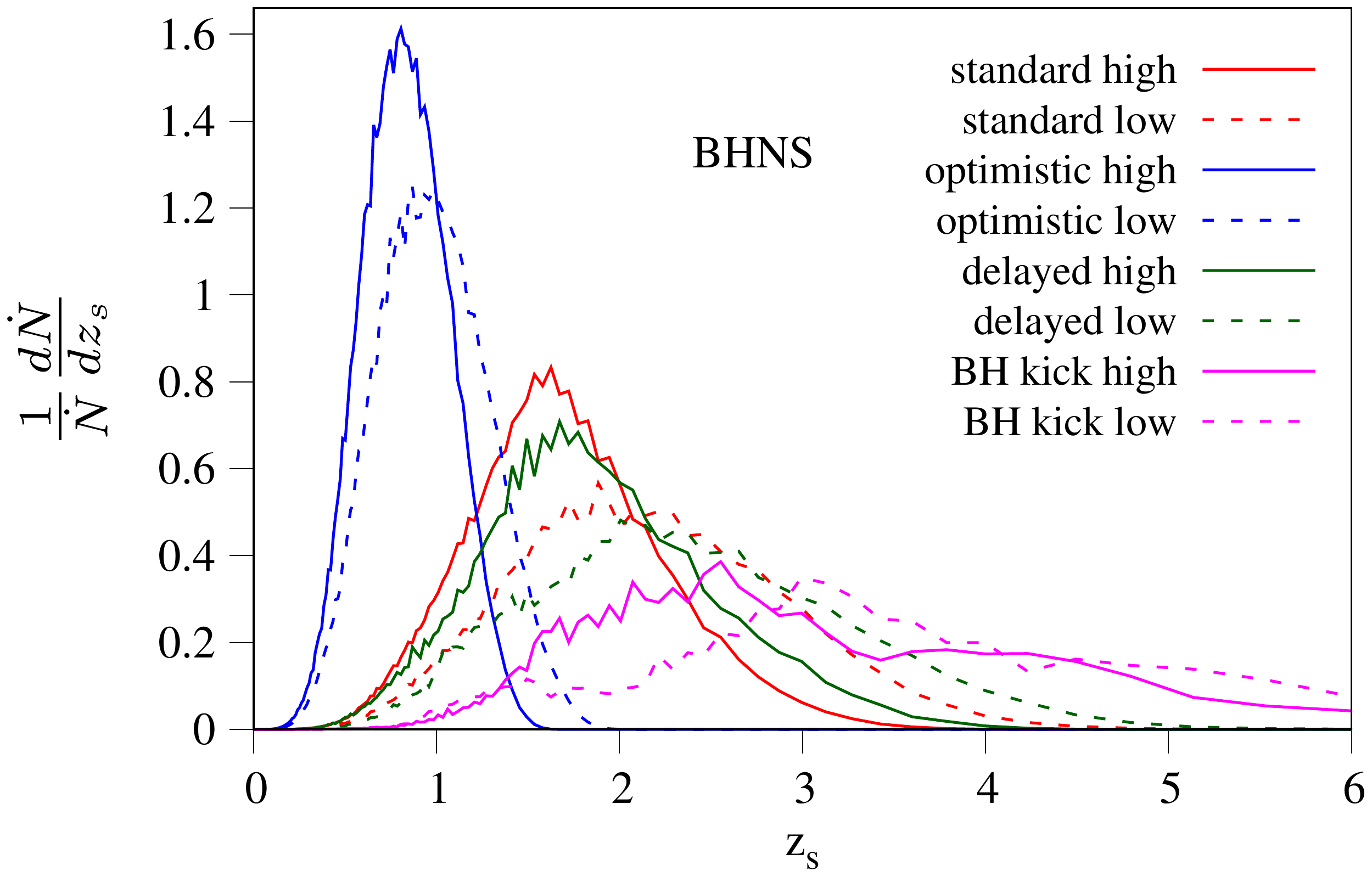}
\includegraphics[width=0.99\columnwidth]{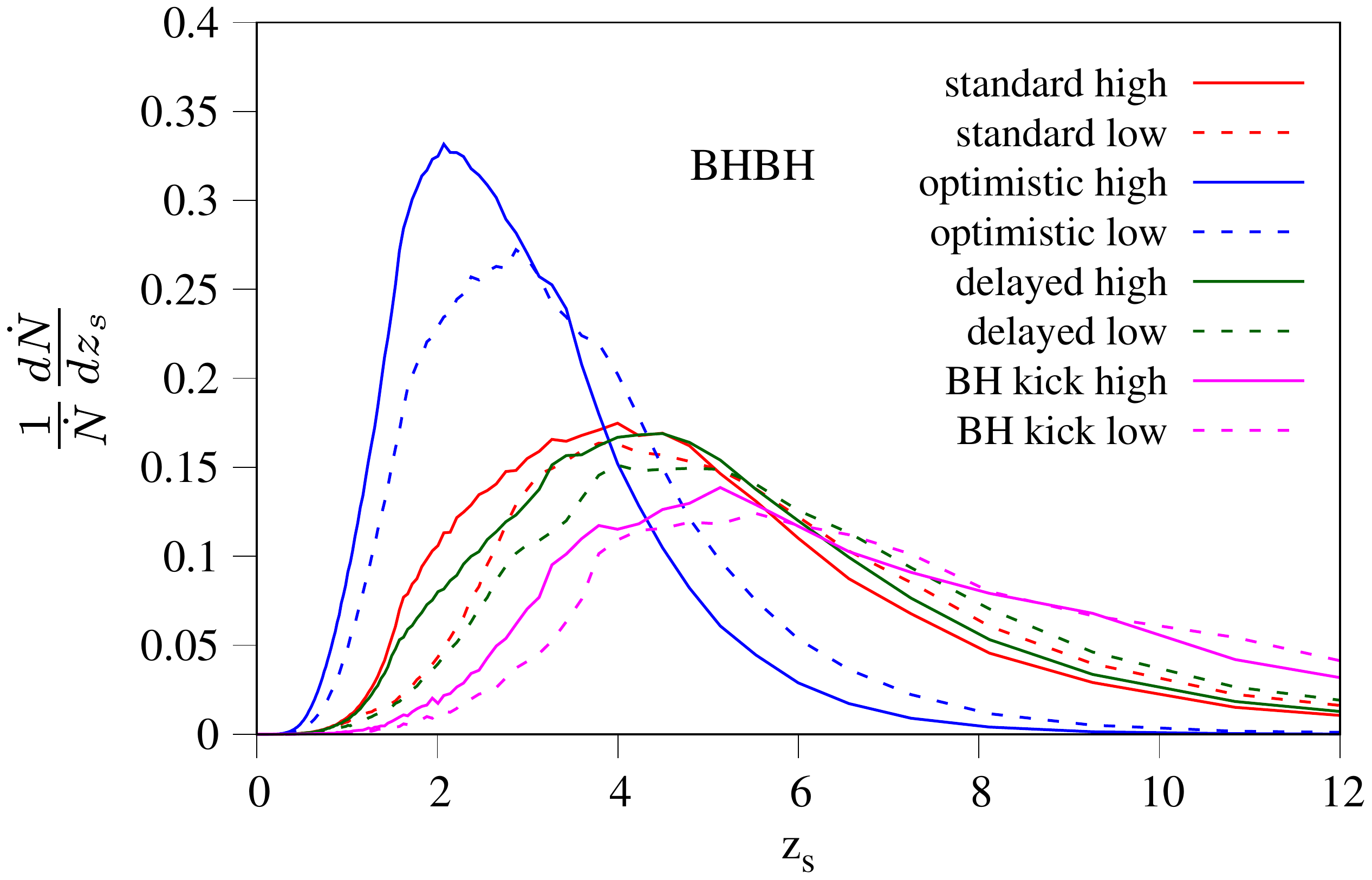}
\caption{Differential lensing rate $ \frac{1}{{\dot N}_{lensed}} \frac{d{{\dot N}_{lensed}}}{dz}$ (as a function of source redshift) of DCO inspiraling binaries for different evolutionary scenarios (solid lines and dashed lines corresponds respectively to high-end and low-end metallicity evolution). Predictions for the DECIGO 4 years operation.
\label{diff lensing}}
\end{figure}

Under the above assumptions, total optical depth for lensing is:
\begin{equation} \label{tau}
\tau_{\pm} = \frac{16}{30} \pi^3 \left( \frac{c}{H_0} \right)^3 {\tilde r}_s^3 \left( \frac{\sigma_{*}}{c} \right)^4 n_{*} \frac{\Gamma \left( \frac{4 + \alpha}{\beta} \right) }{\Gamma \left( \frac{\alpha}{\beta} \right)} y_{\pm, max}^2.
\end{equation}

The above formula is valid only in the case of a continuous search.
If instead the survey has a finite duration $T_{surv}$ some of the
events, i.e. those whose signals come near the beginning or the end
of the survey, would be lost because of lensing time delay. In other
words we would register the signal from just one image and cannot
tell that in fact the event was lensed. Finite duty cycle of the
detector influences the optical depth $\tau_{\pm}$. Namely, it
should be corrected as:
\begin{equation} \label{tauDT}
\tau_{\Delta t,\pm} =  \tau \left[ 1 - \frac{1}{7} \frac{\Gamma \left( \frac{\alpha+8}{\beta} \right)}{\Gamma \left( \frac{\alpha+4}{\beta} \right)} \frac{{\Delta t}_{*, \pm}}{T_{surv}} \right]
\end{equation}
where $\Delta t_{*,\pm} = \frac{32 \pi^2}{H_0} {\tilde r}_s \left(
\frac{\sigma_{*}}{c} \right)^4 y_{\pm, max} $ (for detailed
calculations see \citet{JCAP_ET1}). This correction is particularly
important here, because the DECIGO and B-DECIGO missions are planned
for $T_{surv} = 4\;yr$. We present predictions for the first year of
operation and for the total 4 yrs of nominal duration. 
Moreover, we conservatively assume that the lensed system is
intrinsically loud enough to exceed the detector's threshold $\rho_0
= 8$ and we assume that the fainter image exceeds the threshold.
This means that we only consider $y_{max} = y_{-,max} = 0.5$,
according to Eq.~(\ref{y_condition}).

Cumulative yearly detection of lensed events up to the source redshift $z_s$ can be calculated as:
\begin{equation} \label{lensing_rate}
{\dot N}_{lensed}(z_s) = \int_0^{z_s} \tau_{\Delta t}(z_s, y_{max}, T_{surv}) \frac{d {\dot N}(>\rho_0)}{dz} dz.
\end{equation}

The results for both DECIGO and B-DECIGO are shown in Table~\ref{lensing-hi}. In Fig.~\ref{diff
lensing} normalized differential lensing rate (as a function of
redshift) is displayed. One can verify that DECIGO in its original 
design would be able to register $5 - 6$ strongly lensed NS-NS (or BH-NS) systems, if the contamination from unresolved sources is neglected. 
These systems could be accompanied by the EM counterpart during their final merger phase detectable by ground based
interferometric detectors. Hence, they are the most promising lensed 
sources enabling e.g precise cosmological inference
\citep{NatureComm} or probing dark matter substructure in lensing
galaxies \citep{Liao2018}. However, contamination by unresolved systems (predominantly BH-BH systems) make the chances of detecting lensed NS-NS or BH-NS binaries negligible. On the other hand tens of  lensed inspirals could be seen from the BH-BH binary systems both in DECIGO and B-DECIGO.

\section{Conclusions}

In this paper we have made the predictions of yearly detection rates
of GW signals from the DCO inspirals detectable in the future
decihertz space-borne detectors DECIGO and its smaller scale version
B-DECIGO. All previous papers concerning DECIGO used only analytical
estimates of NS-NS merger rates up to $z=5$. In our calculations, we
have used for the first time the {\tt StarTrack} population
synthesis results concerning intrinsic merger rates at different
redshifts of distinct classes of DCO systems: NS-NS, BH-NS and BH-BH
binaries. Now, when more accurate estimates of merger rates from the LIGO/Virgo data are available \citep{LIGO_run2} one can make a comparison between {\tt StarTrack} predictions and real data estimates 
\citep{beat}. It turns out that ``low-end'' metallicity predictions are in agreement with those based on LIGO/Virgo derived merger rates. 

We have also estimated the stochastic noise levels due to
unresolved DCO systems. The conclusion is that DECIGO would be
significantly contaminated by unresolved BH-BH binary systems, while the level
of this stochastic noise component is not so relevant for B-DECIGO.
Concerning the DCO yearly detection rate its order of magnitude
ranges from $10^2$ for NS-NS, $10^3$ for BH-NS systems to $10^6$ for BH-BH
systems in DECIGO. Respective rates for the B-DECIGO are $10$ (NS-NS), $10^2$ (BH-NS) and  $10^5$ (BH-BH).

The detector's distance parameter $r_0$ for the DECIGO is about 4
times bigger than for the ET, meaning that DECIGO will probe 64
times bigger volume than the next generation of ground based
detectors. Hence, we addressed the issue of gravitationally lensed
DCO inspirals observable by the DECIGO.  For the completness of
discussion we considered B-DECIGO as well, even though B-DECIGO's
$r_0$ is about 3 times smaller than for the ET. Our basic assessment
was performed under assumption that the GW source is intrinsically
loud (i.e. would be detectable without lensing). 
The result is that due to contamination of unresolved systems, either 
DECIGO or B- DECIGO will not be able to register any lensed NS-NS or BH-NS inspirals. However, they could register up to $O(10)$ lensed BH-BH inspirals.

In the appendix we enrich our discussion
relaxing the $\rho_{intr.} \geq 8$ assumption and consider
intrinsically faint signals, i.e. those being detectable exclusively
due to lensing magnification. Such inclusion of lensing
magnification could significantly enlarge the statistics of lensed
events with intrinsic SNR $\rho_{intr.} \geq 8$ for both DECIGO and
B-DECIGO. The appendix also addresses the question of the
magnification bias in the full inspiral GW event catalogs of DECIGO
and B-DECIGO. One finds out that the magnification bias is of the
order of $10^{-3} - 10^{-4}$, which means that the cosmological inferences drawn from these catalogs would not be affected very much.

Let us stress that DECIGO and B-DECIGO would be able to register
inspiral signals from the DCOs weeks or years (the higher redshifted chirp mass the shorter time of passing through DECIGO band) before they enter the high
frequency band of ground-based detectors to finally end their lives
in energetic mergers. Having in mind this circumstance and the
benefits that are expected from registering lensed GW one would expect that DECIGO's
detections of lensed GW signals could trigger a concerted efforts
for searching strong lenses in the EM domain at possible locations
suggested by the DECIGO. This would be very profitable in many
aspects, in particular for identifying the host galaxy before the
merger and despite of no bright EM counterpart as it was the case to
the BH-BH mergers registered so far.  In the light of considerable rate of DCO inspiral signals detectable by the DECIGO, one should be concerned how to distinguish gravitationally lensed signals.   This issue deserves separate studies on simulated mock catalogs of signals. It is usually expected that lensed signals would have the same temporal behaviour (frequency and its rate of change) differing only by amplitude due to magnification and come to the detector after some delay \citep{Dai2017, Tjonnie, JCAP_ET2}. This is a very reasonable expectation for lensed signals from the final merger phase.  
However, unlike the
ground-based detector where one registers a transient event, DECIGO
would observe lensed events in an adiabatic inspiral phase as two 
(or more) unresolved images, likely producing interference patterns
in the waveforms \citep{Hou:2019dcm}. Let us note that interference beat patterns are interesting on their own. This topic is besides the scope of the present paper
which focused only on the statistics of GW lensing and is the subject of a separate study \cite{beat}.

\section*{Acknowledgments} 
We would like to thank the referee for constructive comments which allowed to improve the original version substantially. 
This work was supported by National Key R\&D Program of China No.
2017YFA0402600; the National Natural Science Foundation of China
under Grants Nos. 11690023, 11373014, and 11633001; Beijing Talents
Fund of Organization Department of Beijing Municipal Committee of
the CPC; the Strategic Priority Research Program of the Chinese
Academy of Sciences, Grant No. XDB23000000; the Interdiscipline
Research Funds of Beijing Normal University; and the Opening Project
of Key Laboratory of Computational Astrophysics, National
Astronomical Observatories, Chinese Academy of Sciences.
This work was initiated at Aspen Center for Physics, which
is supported by National Science Foundation grant PHY-1607611. This
work was partially supported by a grant from the Simons Foundation.
M.B. is grateful for this support.
S.H. was supported by Project funded by China Postdoctoral Science Foundation (No.~2020M672400).

\section*{Appendix - Lensing of intrinsically faint GW sources}

Gravitational lensing magnifies the lensed images of the source.
Therefore, one can relax the assumption that sources should be
intrinsically loud enough to be detected without lensing and treat
their signal to noise parameter $\rho$ as a free one. In such a case
one would get estimates for the population of DCO systems detectable
exclusively due to gravitational lensing. In other words this would
provide forecasts for magnification bias in the catalog of lensed
DCO inspirals. We will follow the strategy used in \citet{JCAP_ET3}
in the case of the ET. Therefore, instead of Eq.~(\ref{rate_nl}) we
have to start with the differential inspiral rate per redshift and
per SNR parameter $\rho$:
\begin{equation} \label{diff_rate}
\frac{\partial^2 {\dot N}}{\partial z_s \partial \rho} = 4 \pi \left( \frac{c}{H_0} \right)^3 \frac{{\dot n_0}(z_s)}{1+z_s} \frac{{\tilde r}^2(z_s)}{E(z_s)} P_{\Theta} (x(z_s, \rho)) \frac{x(z_s, \rho)}{\rho}
\end{equation}
then one should calculate cross-sections $S_{cr, \pm}$  and optical
depths $\tau_{\pm}$ for lensing separately for each image $I_{+}$ or
$I_{-}$. 

\begin{figure}[htbp]
    \centering
    \includegraphics[width=0.92\columnwidth]{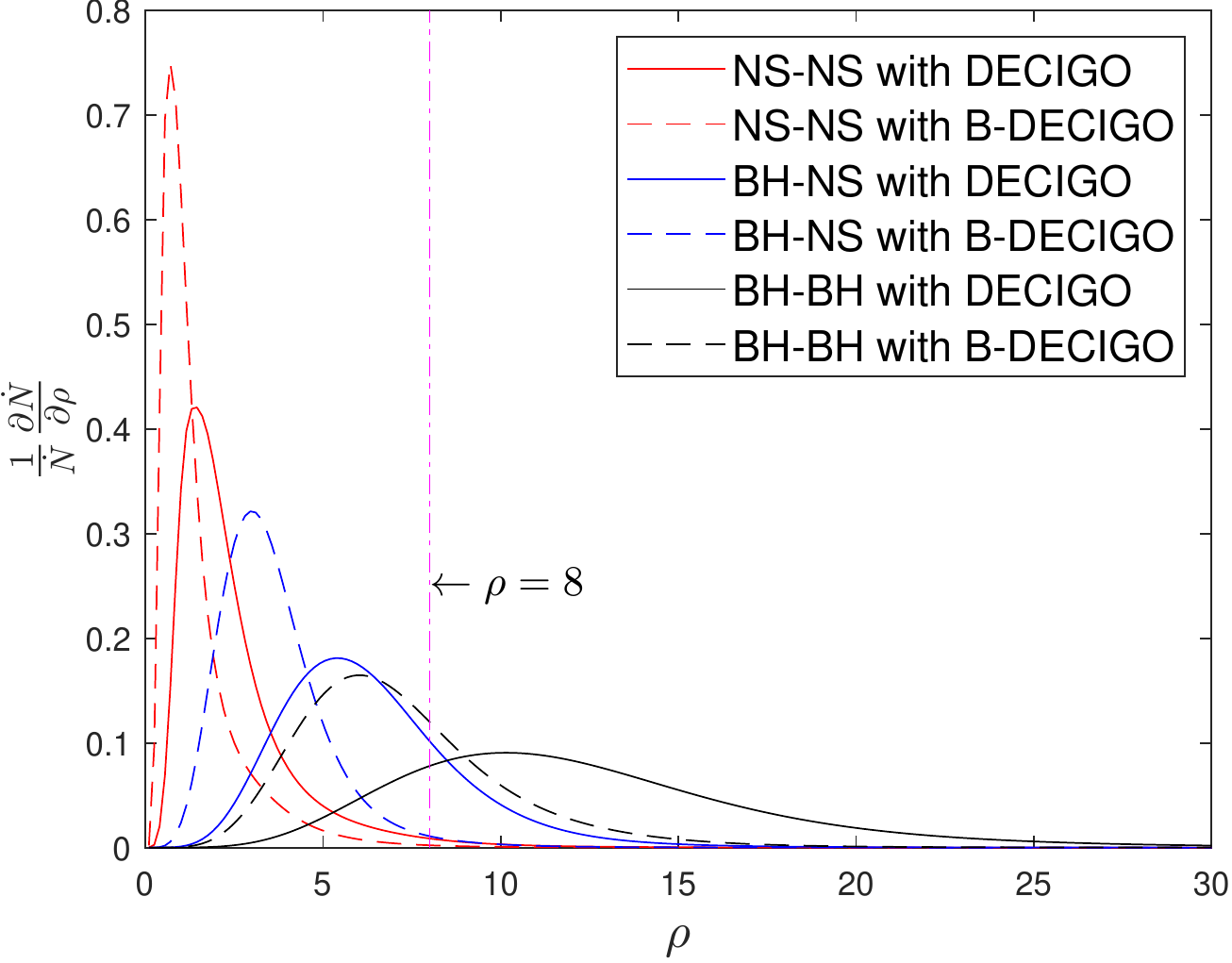}
    \caption{The normalized differential yearly detection rates $\frac{1}{\dot N}\frac{\partial\dot N}{\partial \rho}$ v.s. the intrinsic SNR $\rho$.
    ``Low-end'' metallicity galaxy evolution and the standard model of DCO formation are assumed. }
    \label{fig-1-ding}
\end{figure}

\begin{figure}[htbp]
\centering
    \includegraphics[width=0.92\columnwidth]{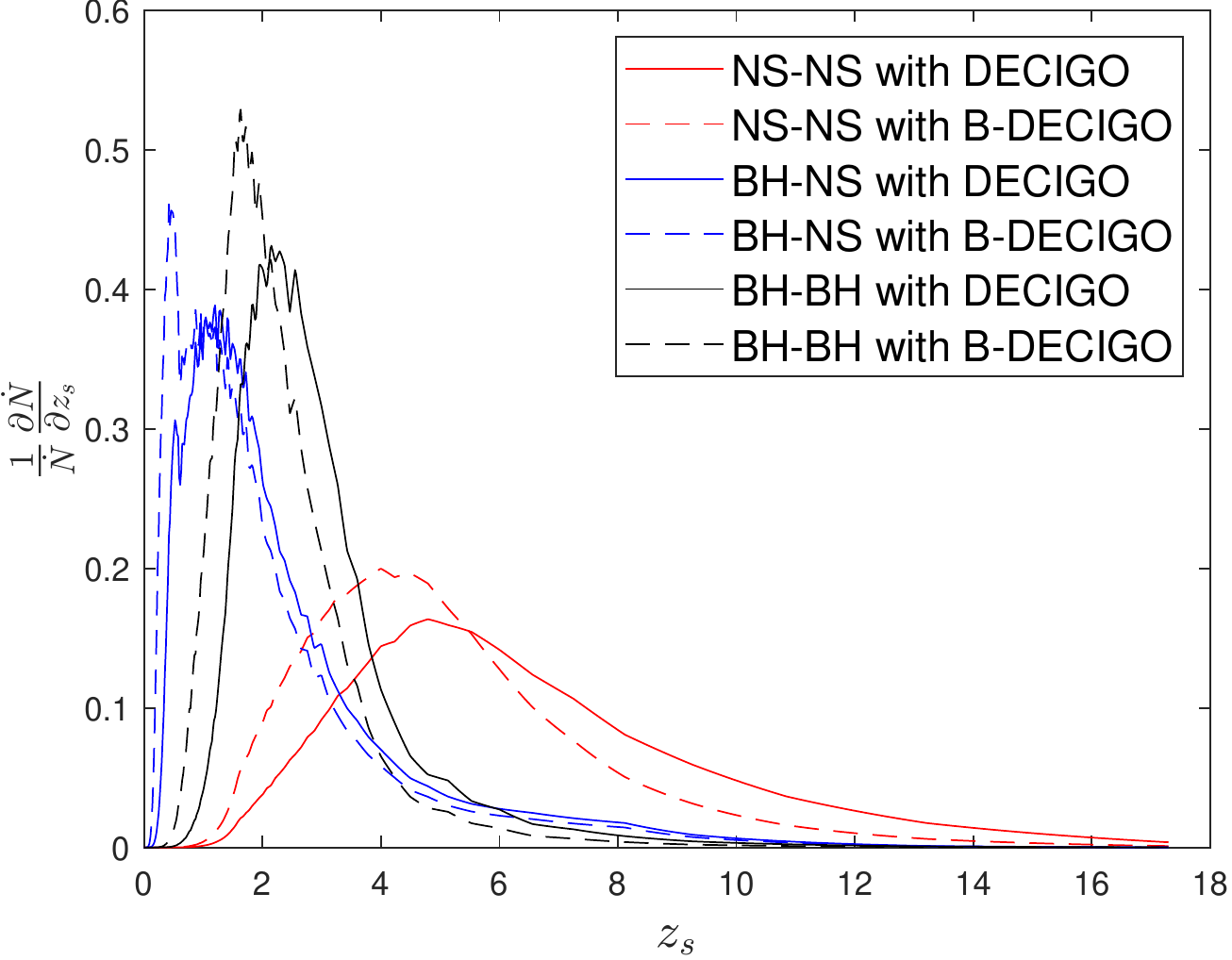}
    \includegraphics[width=0.92\columnwidth]{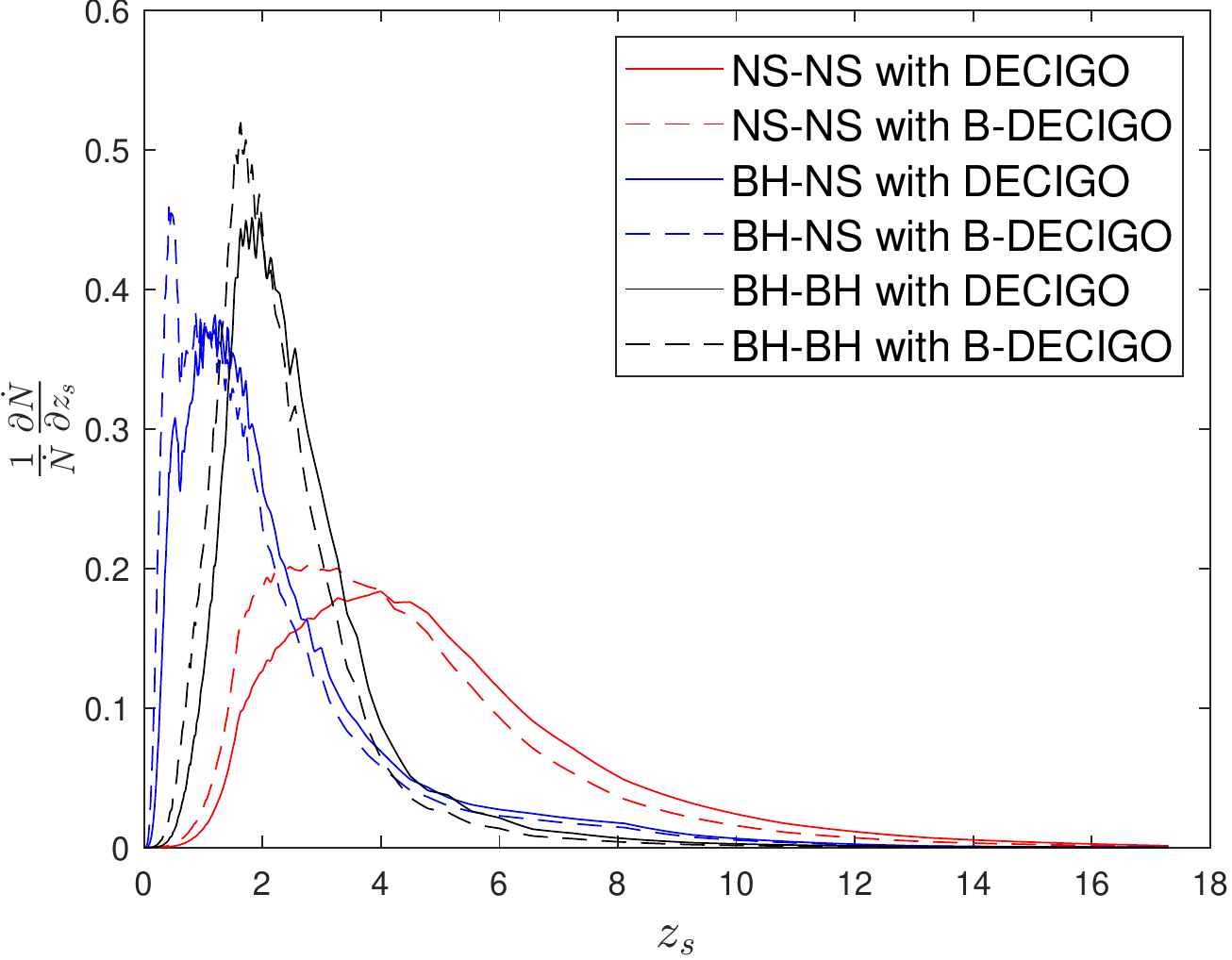}
    \caption{The normalized differential yearly detection rates $\frac{1}{\dot N}\frac{\partial\dot N}{\partial z_s}$ v.s. the source redshift $z_s$. Upper figure corresponds to $\rho_{intr} < 8$ for the $I_-$ image. The lower one corresponds to the $I_-$ image including both $\rho_{intr} < 8$  and $\rho_{intr} \geq 8$ .  Low-end metallicity galaxy evolution  and standard model of DCO formation are assumed. 
    }
    \label{fig-2-ding}
\end{figure}

\begin{table*}[htbp]
\footnotesize 
\caption{Expected numbers of lensed GW events from inspiraling DCOs
of different classes under different evolutionary scenarios,
assuming ``low-end'' and ``high-end'' metallicity evolution.
Predictions for the DECIGO \color{black}{and B-DECIGO} under assumption of $T_{surv}=4\;yrs$.  } \label{lensing-hi}
\begin{center}  
\begin{tabular}{cccccc}
\hline
\\
&& DECIGO &&\\
\\
\hline \\
Evolutionary scenario  & standard & optimistic CE & delayed SN& high BH kicks  \\
\\
\hline
NS-NS & & & & \\
low-end metallicity &0.&0.&0.&0.07\\
high-end metallicity &0.&0.&0.&0.29\\

\hline
BH-NS & & & & \\
low-end metallicity & 0.2 &	0.02 & 0.15 & 0.38 \\
high-end metallicity & 0.21 & 0.03 & 0.2 & 0.39 \\

\hline
BH-BH & & & & \\
low-end metallicity & 66.91 & 58.12 & 62.86 & 10.04 \\
high-end metallicity & 65.07 & 71.28 & 61.41 & 8.46 \\

\hline 

\\
&& B-DECIGO &&\\
\\
\hline \\
Evolutionary scenario  & standard & optimistic CE & delayed SN& high BH kicks  \\
\\

\hline
NS-NS & & & & \\
low-end metallicity & 0. & 0. & 0. &  0.\\
high-end metallicity & 0. & 0. &  0. & 0.\\

\hline
BH-NS & & & & \\
low-end metallicity & 0.2 & 0.02 & 0.15 & 0.38 \\
high-end metallicity &  0.21 & 0.03 & 0.2 & 0.39 \\
\hline
BH-BH & & & & \\
low-end metallicity &  9.25 & 5.42 & 9.2 & 2.73 \\
high-end metallicity &  13.66 & 10.94 & 14.78 &	2.25 \\

\hline

\end{tabular}
\end{center}
\end{table*}

\begin{table*}[htbp]
\footnotesize 
\caption{Expected numbers of lensed GW events observed by DECIGO and B-DECIGO
with $\rho_{intr}<8$ for which the $I_-$ image is magnified above
threshold $\rho_0 = 8$. We assumed the survey duration $T_{surv}=4\;yrs$. Nomenclature of DCO formation scenarios and galaxy metallicity
evolution follows that of \citet{BelczynskiII}. 
} \label{lensing-1}
\begin{center}  
\begin{tabular}{cccccc}

\hline
\\
&& \color{black}{DECIGO} &&\\
\\

\hline

Evolutionary scenario  & standard & optimistic CE & delayed SN & high BH kicks  \\
\hline

NS-NS & & & & \\
low-end metallicity &  0.02 & 0.03 &  0.03 &  0.64 \\
high-end metallicity & 0.06 & 0.05 & 0.10 & 0.98 \\

\hline

BH-NS & & & & \\
low-end metallicity &  0.83 &  0.38 & 0.49 &  0.13 \\
high-end metallicity & 0.86 & 0.48 & 0.52 & 0.12 \\

\hline

BH-BH & & & & \\
low-end metallicity &  26.6 &  66.1 &  21.5 &  0.62 \\
high-end metallicity & 21.1 & 66.1 & 16.3 & 0.45 \\

\hline

TOTAL & & & & \\
low-end metallicity &  27.5 &  66.5 &  22.0 &  1.39 \\
high-end metallicity & 22.0 & 66.6 & 16.9 & 1.55 \\

\hline
\\
&& \color{black}{B-DECIGO} &&\\
\\
\hline

Evolutionary scenario  & standard & optimistic CE & delayed SN & high BH kicks  \\
\hline

NS-NS & & & & \\
low-end metallicity &  0.002 & 0.003 &  0.002 &  0.01 \\
high-end metallicity & 0.004 & 0.004 & 0.005 & 0.02 \\

\hline

BH-NS & & & & \\
low-end metallicity &  0.122 &  0.054 & 0.08 &  0.06 \\
high-end metallicity & 0.112 & 0.064 & 0.08 & 0.04 \\

\hline

BH-BH & & & & \\
low-end metallicity &  24.3 &  31.5 &  22.0 &  2.39 \\
high-end metallicity & 22.4 & 34. & 20.1 & 1.98 \\

\hline

TOTAL & & & & \\
low-end metallicity &  24.4 &  31.6 &  22.1 &  2.5 \\
high-end metallicity & 22.5 & 34.6 & 20.2 & 2.0 \\

\hline
\end{tabular}
\end{center}
\end{table*}

The final result will be the lensing rate of intrinsically
faint ($\rho < \rho_0$) DCOs having $I_{+}$ or $I_{-}$ images
magnified above the threshold $\rho_0$:
\begin{equation} \label{lensing_rate}
{\dot N}_{lensed, \pm} = \int_0^{z_{max}} dz_s \int_0^{\rho_0} \tau_{\Delta t,\pm}(z_s, \rho) \frac{\partial^2 {\dot N}}{\partial z_s \partial \rho} d \rho
\end{equation}
or differential lensing rates with respect to $\rho$ or $z_s$
respectively:
\begin{equation} \label{diff_lensing_rate_rho}
\frac{d {\dot N}_{lensed, \pm}}{d \rho} = \int_0^{z_{max}} \tau_{\Delta t,\pm}(z_s, \rho) \frac{\partial^2 {\dot N}}{\partial z_s \partial \rho} dz_s
\end{equation}
\begin{equation} \label{diff_lensing_rate_zs}
\frac{d {\dot N}_{lensed, \pm}}{d z_s} = \int_0^{\rho_0}  \tau_{\Delta t,\pm}(z_s, \rho) \frac{\partial^2 {\dot N}}{\partial z_s \partial \rho} d \rho.
\end{equation}

\begin{table*}[htbp]
\footnotesize 
\caption{Expected numbers of lensed GW events observed by DECIGO and B-DECIGO
with $\rho_{intr}<8$ for which the $I_+$ image is magnified above
threshold $\rho_0 = 8$. Other assumptions and terminology -- like in
Table~\ref{lensing-1}. } \label{lensing-2}
\begin{center}  
\begin{tabular}{cccccc}

\hline
\\
&& \color{black}{DECIGO} &&\\
\\

\hline
\\
Evolutionary scenario  &standard & optimistic CE & delayed SN & high BH kicks  \\
\\
\hline

NS-NS & & & & \\
low-end metallicity &  0.04 &  0.04 &  0.067 &  2.31 \\
high-end metallicity & 0.14 & 0.061 & 0.254 & 4.08 \\

\hline

BH-NS & & & & \\
low-end metallicity &  3.94 &  1.30 &  2.43 &  0.68\\
high-end metallicity & 4.16 & 1.72 & 2.60 & 0.630\\

\hline

BH-BH & & & & \\
low-end metallicity &  135.6 & 335.7 &  109.4 &  3.11 \\
high-end metallicity & 107.4 & 333.0 & 82.6 & 2.29 \\

\hline

TOTAL & & & & \\
low-end metallicity & 139.6 &  348.7 &  111.9 &  6.1\\
high-end metallicity & 111.7 & 334.8 & 85.4 & 7.0\\

\hline
\\
&& \color{black}{B-DECIGO} &&\\
\\
\hline
\\
Evolutionary scenario  &standard & optimistic CE & delayed SN & high BH kicks  \\
\\
\hline

NS-NS & & & & \\
low-end metallicity &  0.002 &  0.003 &  0.003 &  0.017 \\
high-end metallicity & 0.005 & 0.005 & 0.008 & 0.032 \\

\hline

BH-NS & & & & \\
low-end metallicity &  0.33 &  0.11 &  0.22 &  0.23\\
high-end metallicity & 0.29 & 0.13 & 0.22 & 0.17\\

\hline

BH-BH & & & & \\
low-end metallicity &  120.0 & 141.0 &  109.4 &  12.1 \\
high-end metallicity & 111.4 & 158.1 & 100.7 & 10.0 \\

\hline

TOTAL & & & & \\
low-end metallicity & 120.3 &  141.1 &  109.6 &  12.3\\
high-end metallicity & 111.7 & 158.2 & 100.9 & 10.2\\

\hline
\end{tabular}
\end{center}
\end{table*}

\begin{figure*}[htbp]
    \centering
    \includegraphics[width=0.92\columnwidth]{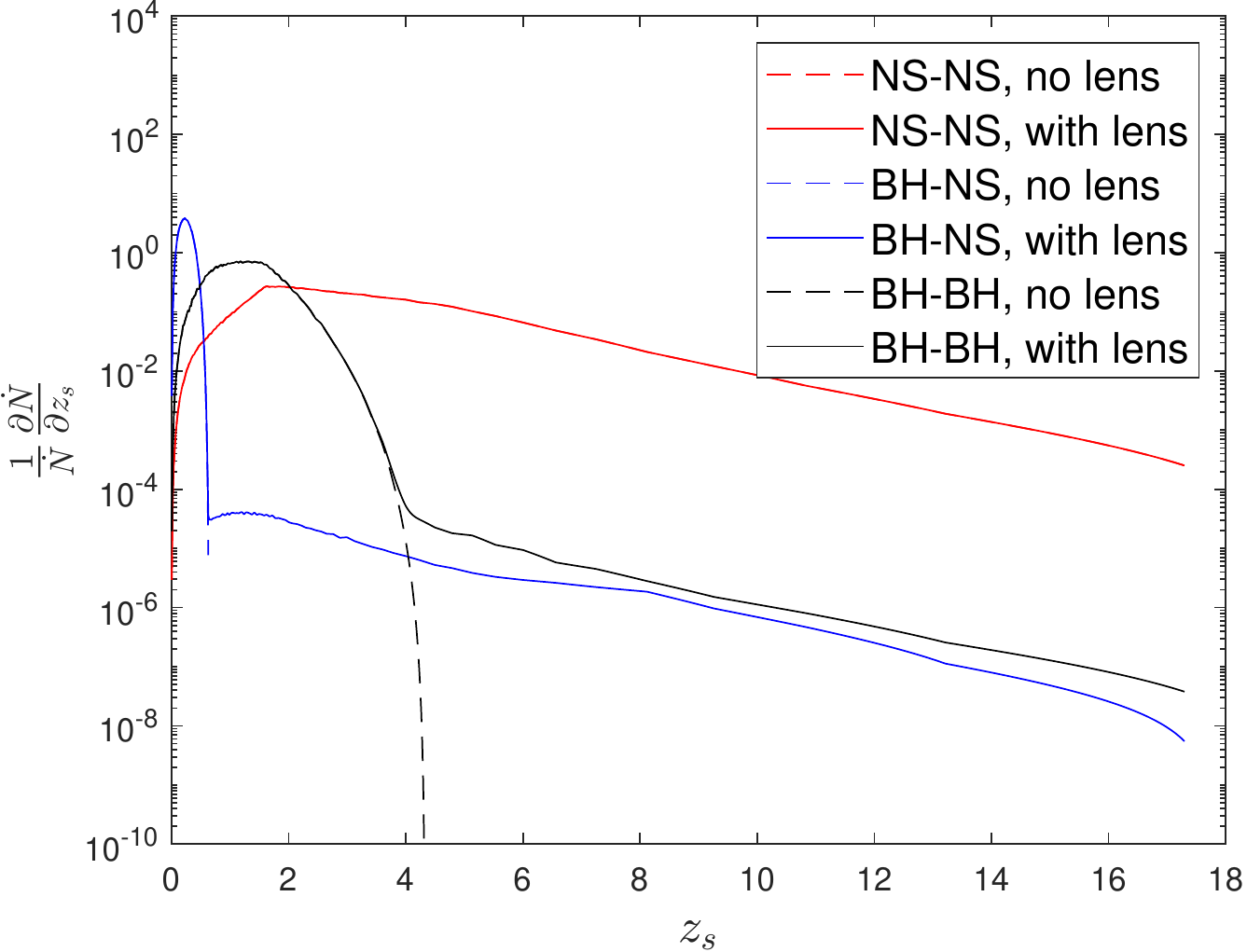}\quad\quad
    \includegraphics[width=0.92\columnwidth]{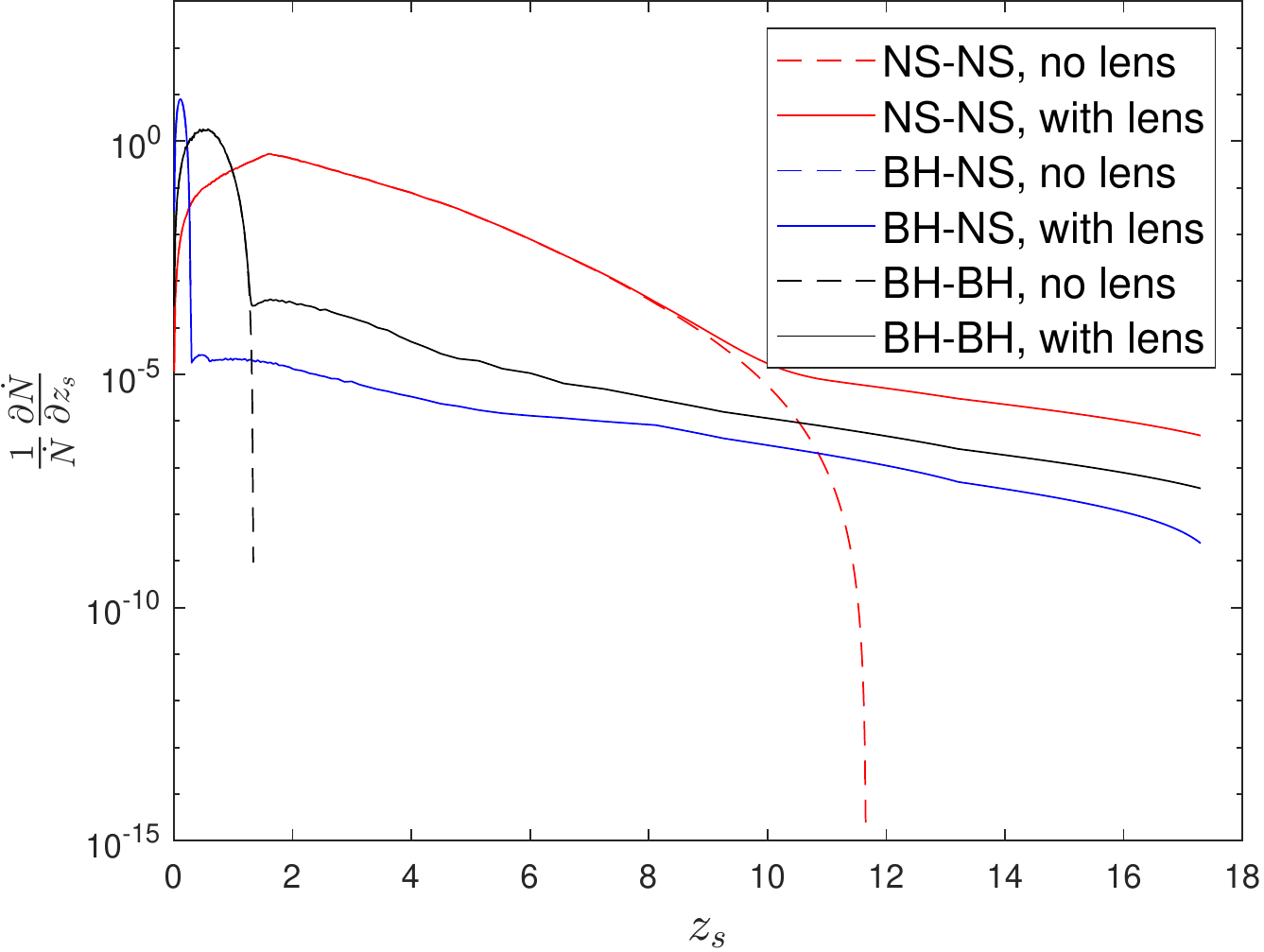}
    \caption{Probability densities of DCO inspiral events yearly rate to be detected by DECIGO (left panel) and B-DECIGO (bright  panel) during their survey duty cycles $T_{surv}=4$ years.
    The solid curves are for the total catalogue of lensed and non-lensed systems, and the dashed curves are for the non-lensed ones.
    Note the logarithmic scale used in this figure.
    }
    \label{fig-3-dec-bdec}
\end{figure*}

Table~\ref{lensing-1} contains the results of predicted
intrinsically faint lensed GW events for which the the $I_-$ image
is magnified above threshold $\rho_0 = 8$ for DECIGO and B-DECIGO. As one can see, inclusion
of lensing magnification of intrinsically faint events would
significantly enlarge the statistics of lensed events with intrinsic
SNR $\rho_{intr} \geq 8$ (see Tab.~\ref{lensing-hi}). This conclusion is true both for
DECIGO and B-DECIGO.

The normalized differential yearly detection rates $\frac{1}{\dot
N}\frac{\partial\dot N}{\partial \rho}$ of lensed events as
functions of the intrinsic SNR $\rho$ for various types of DCOs to
be observed by DECIGO and B-DECIGO are displayed in
Fig.~\ref{fig-1-ding}.

Figure \ref{fig-2-ding} shows the normalized differential yearly
detection rates $\frac{1}{\dot N}\frac{\partial\dot N}{\partial
z_s}$ of lensed events as functions of the source redshift $z_s$.
The top panel illustrates lensed faint systems with $\rho_{intr}<8$
with the $I_-$ image magnified above the threshold. In other words,
both images will be registered by DECIGO or B-DECIGO. In the lower
panel differential detection rate is shown for systems of both
$\rho_{intr}<8$ and $\rho_{intr} \geq 8$, i.e. for the total catalog
of lensed GW events of  DECIGO or B-DECIGO. For the sake of
transparency only standard scenario of DCO formation with
``low-end'' metallicity evolution is shown. Detector's operation
period of $T_{surv}=4$ years is assumed. From these figures, one
infers that the faint sources can be used to probe higher redshifts,
and these higher redshift sources would thus contaminate the future
catalog of gravitationally lensed GW events. This is expected on the
ground of general idea of how the magnification bias works. Since
DECIGO's characteristic radius $r_0$ is much larger than that of
B-DECIGO, the former configuration is more suitable to probe high
redshift sources, as shown by Fig.~\ref{fig-2-ding}.

Besides the magnification bias on the lensed GW events, the
magnification bias at the level of the full DCO inspiral events
catalogue can also be obtained. For this end, one may calculate the
detection rate of the intrinsically faint events whose $I_+$ image
is magnified above the threshold. Table~\ref{lensing-2} shows the
predictions for DECIGO and B-DECIGO. One can clearly see the increase in the detection rates,
since $y_{+,max}>y_{-,max}$. 
Comparing this table with
Table~\ref{rates}, one finds out that the
magnification bias at the level of the resolvable inspiral DECIGO or B-DECIGO event catalogs
would be of the order of $10^{-3} - 10^{-4}$  depending on the DCO
population. This means that the cosmological inferences drawn from
this catalog would not be affected very much. 

One can demonstrate
this effect by plotting together probability density of yearly
detection rate of non-lensed sources and total prediction, which is
shown in Fig.~\ref{fig-3-dec-bdec} for DECIGO and
B-DECIGO. The upper panel of Fig.~\ref{fig-3-dec-bdec} shows that the
magnification bias is negligible for all three types of the DCO
sources in the case of DECIGO. For B-DECIGO, the magnification bias
is negligibly small for NS-NS and BH-NS binaries, while for BH-BH
binaries, it is barely noticeable, according to the lower panel of
Fig.~\ref{fig-3-dec-bdec}.


\end{document}